# Analytical Estimation in Differential Optical Transmission Systems Influenced by Equalization Enhanced Phase Noise

Tianhua Xu[1,*], Gunnar Jacobsen[2,3], Sergei Popov[2],
Tiegen Liu[4], Yimo Zhang[4], and Polina Bayvel[1]

1. University College London, London, WC1E7JE, United Kingdom
2. Royal Institute of Technology, Stockholm, SE-16440, Sweden
3. Acreo Swedish ICT AB, Stockholm, SE-16440, Sweden
4. Tianjin University, Tianjin, 300072, China

(*Invited Paper*)

**Abstract**— An analytical model is presented for assessing the performance of the bit-error-rate (BER) in the differential *m*-level phase shift keying (*m*-PSK) transmission systems, where the influence of equalization enhanced phase noise (EEPN) has been considered. Theoretical analysis has been carried out in differential quadrature phase shift keying (DQPSK), differential 8-PSK (D8PSK), and differential 16-PSK (D16PSK) coherent optical transmission systems. The influence of EEPN on the BER performance, in term of signal-to-noise ratio (SNR), are investigated for different fiber dispersion, LO laser linewidths, symbol rates, and modulation formats. Our analytical model achieves a good agreement with previously reported EEPN induced BER floors, and can give an accurate prediction for the DQPSK system, and a leading-order approximation for the D8PSK and the D16PSK systems.

## 1. INTRODUCTION

The performance of long-haul high-speed optical fiber communication systems is severely degraded by the impairments in the transmission channels and laser sources, such as chromatic dispersion (CD), polarization mode dispersion (PMD), laser phase noise (PN) and fiber nonlinear effects [1-6]. Coherent optical detection and digital signal processing (DSP) allow the powerful equalization and effective mitigation of the system impairments in the electrical domain [7-12], and have become one of the most promising techniques for the next-generation optical fiber communication networks to achieve a performance close to the Shannon capacity limit [13,14]. Using high-level modulation formats such as the *m*-level phase shift keying (*m*-PSK), the performance of optical fiber transmission systems will be degraded seriously by the phase noise from the transmitter (Tx) lasers and the local oscillator (LO) lasers [15,16]. It has been demonstrated that the laser phase noise can be effectively compensated using the feedforward and the feedback carrier phase recovery (CPR) methods [17-22]. As a conventional feedforward approach, the differential carrier phase recovery has been validated as an effective method for the laser phase noise compensation in coherent communication systems, and can be regarded as a benchmark for evaluating other carrier phase recovery methods [22].

In the electronic dispersion compensation (EDC) based coherent optical fiber transmission systems, an effect of equalization enhanced phase noise (EEPN) is generated due to the interactions between the EDC module and the LO laser phase noise (in the post-EDC case) [23-26]. The performance of long-haul optical fiber communication systems will be degraded significantly due to the EEPN, with the increment of fiber dispersion, LO laser linewidths, modulation levels, symbol rates and system bandwidths [27-29]. The impacts of EEPN have been investigated in the single-channel, the WDM multi-channel, the orthogonal frequency division multiplexing (OFDM), the dispersion pre-distorted, and the multi-mode optical fiber communication systems [29-33]. Meanwhile, some studies have also been carried out to investigate the influence of EEPN in the differential carrier phase recovery in long-haul high-speed optical communication systems [28].

In this paper, an analytical model is presented for assessing the performance of the bit-error-rate (BER) in the differential *m*-PSK transmission systems, where the influence of EEPN is considered. Theoretical analysis has been carried out in differential quadrature phase shift keying (DQPSK), differential 8-PSK (D8PSK), and differential 16-PSK (D16PSK) optical transmission systems. The influence of EEPN on the BER performance in term of signal-to-noise ratio (SNR), are analyzed for different fiber dispersion, LO laser linewidths, symbol rates, and modulation formats. Our analytical model achieves a good agreement with previously reported EEPN induced BER floors, and can give an accurate prediction for the DQPSK system, and a leading-order approximation for the D8PSK and the D16PSK systems. Our analytical model can also be applied for evaluating the BER performance in the coherent *m*-PSK transmission systems employing the one-tap normalized least-mean-square carrier phase recovery.

*Corresponding author: Tianhua Xu (tianhua.xu@ucl.ac.uk)

## 2. THEORY OF EEPN AND DIFFERENTIAL CARRIER PHASE RECOVERY

In the coherent optical communication systems employing post EDC and differential CPR, the Tx laser phase noise passes through both the optical fiber and the EDC module, and its net experienced dispersion is close to zero. However, the LO phase noise only goes through the EDC module, of which the transfer function is heavily dispersed in the systems without using optical dispersion compensation. Thus the LO phase noise will interact with the EDC module, and will significantly affect the performance of the long-haul high speed coherent transmission systems [23,24].

It has been demonstrated that the EEPN variance scales linearly with the accumulated CD, the LO laser linewidth, and the symbol rate [10,23]. The variance of the additional noise due to the EEPN can be described as

$$\sigma^2_{EEPN} = \pi\lambda^2 \cdot D \cdot L \cdot \Delta f_{LO} / 2cT_S \qquad (1)$$

where $\lambda$ is the central wavelength of the optical carrier, $c$ is the light speed in vacuum, $D$ is the CD coefficient of the transmission fiber, $L$ is the fiber length, $\Delta f_{LO}$ is the 3-dB linewidth of the LO laser, and $T_S$ is the symbol period of the transmission system.

In the differential carrier phase recovery, the carrier phase is recovered based on the phase difference between two consecutive received symbols [22]. When the EEPN is taken into consideration, the performance of BER versus SNR in the differential *m*-PSK coherent optical transmission systems can be expressed as

$$BER^{EEPN}_{m-PSK} \approx \frac{4}{\sqrt{2\pi}m\sigma \log_2 m} \int_{-\infty}^{+\infty} \exp\left(\frac{-8\varepsilon^2}{m^2\sigma^2}\right) \cdot erfc\left[\left(\sqrt{1+\sin\frac{\pi}{m}} - \sqrt{1-\sin\frac{\pi}{m}}\right) \cdot \sqrt{SNR}\right] d\varepsilon \qquad (2)$$

$$\sigma^2 \approx \sigma^2_{Tx} + \sigma^2_{LO} + \sigma^2_{EEPN} + 2\rho \cdot \sigma_{LO}\sigma_{EEPN} \approx 2\pi\Delta f_{Tx} \cdot T_S + 2\pi\Delta f_{LO} \cdot T_S + \frac{\pi\lambda^2}{2c} \cdot \frac{D \cdot L \cdot \Delta f_{LO}}{T_S} \qquad (3)$$

where *m* is the modulation level, $\sigma^2$ is the total noise variance, $\sigma^2_{Tx}$ and $\sigma^2_{LO}$ are the phase noise variance of the Tx laser and the LO laser respectively, $\Delta f_{Tx}$ is the 3-dB linewidth of the Tx laser, and $\rho$ is the correlation coefficient between the LO phase noise and the EEPN.

## 3. RESULTS AND DISCUSSIONS

Based on the proposed analytical model, the behavior of the BER versus SNR in the optical transmission system has been investigated by considering the influence of EEPN. The performance of optical communication systems using differential carrier phase recovery was studied in terms of different laser linewidths, different modulation formats and different symbol rates. In all these analyses, the CD coefficient of the transmission fiber is 16 ps/nm/km, and the attenuation, the PMD, and the fiber nonlinearities are neglected.

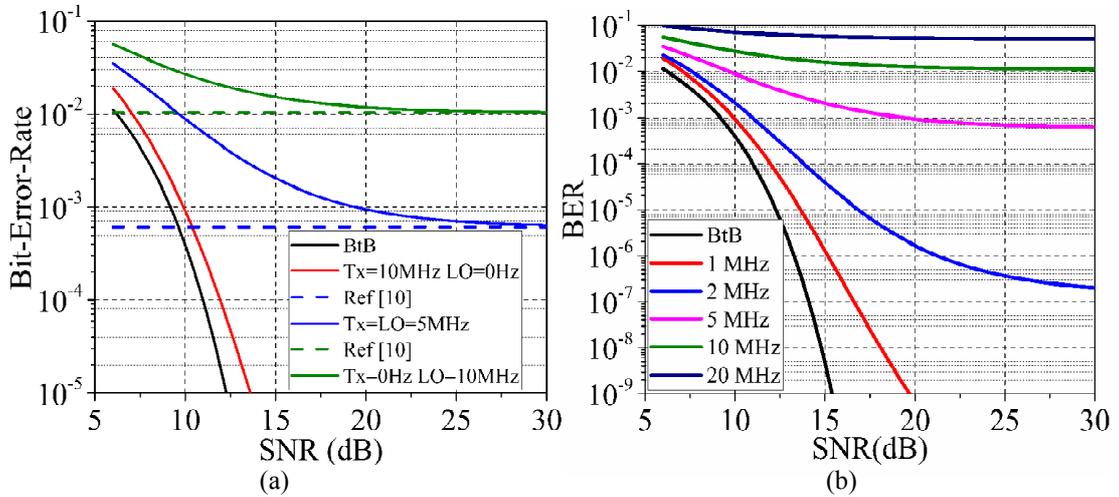

Figure 1. Performance of BER versus SNR in the 28-Gbaud DQPSK optical transmission systems. (a) different distributions of the Tx and the LO laser linewidths, (b) different LO laser linewidths (Tx laser linewidth is equal to LO laser linewidth).

As shown in Fig. 1, the performance of the 28-Gbaud DQPSK optical communication system over 2000 km standard single-mode fiber (SSMF) transmission has been investigated based on the proposed analytical model. The results in Fig. 1(a) were studied by considering the equalization enhanced phase noise under different distributions of the Tx laser and the LO laser linewidths. It can be found that the EEPN originates from the LO laser phase noise rather than the Tx phase noise in the transmission systems using post-EDC, and that the limits of the BER achieve a very good agreement with the BER floors (the dash curve) reported in our previous work [10]. The performance of the 28-Gbaud DQPSK transmission system has also been investigated in terms of different LO laser linewidths, as shown in Fig. 1(b), where the Tx laser linewidth is set equal to the LO laser linewidth. It can be seen that the performance of the BER in the differential transmission systems degrades significantly with the increment of the LO laser linewidths.

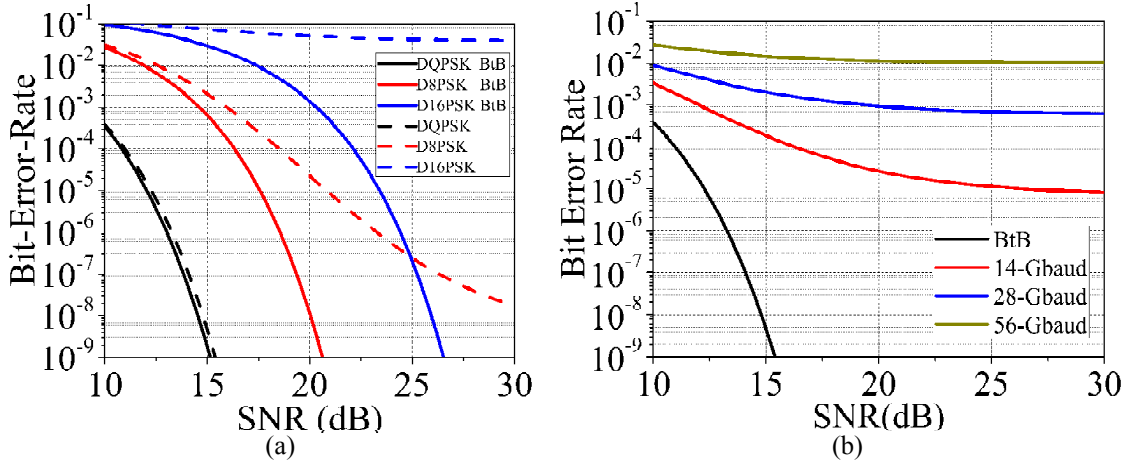

Figure 2. Performance of BER versus SNR in differential optical transmission systems using different modulation formats and different symbol rates. (a) different modulation formats at 28-Gbaud, (b) different symbol rates for DQPSK system.

The performance of BER versus SNR in the coherent optical transmission system using different modulation formats is shown in Fig. 2(a), where the DQPSK, the D8PSK, and the D16PSK transmission systems are investigated at 28-Gbaud. The linewidths of the Tx and the LO lasers are both set as 100 kHz. A transmission distance of 2000 km has been applied to study the impact of the EEPN (rather than the intrinsic laser phase noise). It is found that the BER performance is degraded by EEPN more severely with the increment of modulation format levels, and the D16PSK system is the most sensitive to the EEPN. In Fig. 2(b), the behaviors of BER versus SNR in the 2000 km DQPSK coherent transmission system are studied using different symbol rates, where both the Tx and the LO lasers linewidths are 5 MHz. Three symbol rates of 14-Gbaud, 28-Gbaud, and 56-Gbaud are considered for the evaluation. We find that the BER behavior is degraded significantly due to a severer EEPN with the increment of symbol rates. Consequently, the requirement of laser linewidths cannot be relaxed in the long-haul transmission systems with a higher symbol rate, due to the influence of EEPN.

4. CONCLUSION

In this paper, the analytical model is proposed for evaluating the performance of BER versus SNR in differential $m$-PSK transmission systems, considering the influence of EEPN. The impact of EEPN are investigated in terms of different fiber dispersion, LO laser linewidths, symbol rates, and modulation formats. The results from this theoretical model achieve a good agreement with the previously reported BER floors. This analytical model can be applied for predicting the BER behaviors in the differential carrier phase recovery and the one-tap normalized least-mean-square feedback carrier phase recovery, and can also be employed as a reference for evaluating other CPR methods in the long-haul high speed coherent optical transmission systems.

ACKNOWLEDGEMENT
This work is supported in parts by UK EPSRC project UNLOC (EP/J017582/1), EU project GRIFFON (No. 324391), and EU project ICONE (No. 608099).